# Numerical Simulations of Particulate Suspensions via a Discretized Boltzmann Equation

# Part I. Theoretical Foundation

Anthony J. C. Ladd

*Lawrence Livermore National Laboratory,*

*Livermore, California 94550.*

(June 25, 1993)

## Abstract

A new and very general technique for simulating solid-fluid suspensions is described; its most important feature is that the computational cost scales *linearly* with the number of particles. The method combines Newtonian dynamics of the solid particles with a discretized Boltzmann equation for the fluid phase; the many-body hydrodynamic interactions are fully accounted for, both in the creeping-flow regime and at higher Reynolds numbers. Brownian motion of the solid particles arises spontaneously from stochastic fluctuations in the *fluid* stress tensor, rather than from random forces or displacements applied directly to the particles. In this paper, the theoretical foundations of the technique are laid out, illustrated by simple analytical and numerical examples; in the companion paper, extensive numerical tests of the method, for stationary flows, time-dependent flows, and finite Reynolds number flows, are reported.

# I. INTRODUCTION

Numerical simulations, which take explicit account of the hydrodynamic forces between the suspended particles, are becoming useful tools for studying the dynamical and rheological properties of suspensions. There are at least three important flow regimes which can be addressed by numerical simulation: colloidal suspensions of sub-micron sized particles, where Brownian forces and viscous forces balance; suspensions of macroscopic particles (*i.e.* larger than $10\mu m$), where the viscous forces alone are important; and flows at small but non-zero Reynolds number ($1 < R_e < 100$). At present, computational cost is the limiting factor; even with supercomputers, it is not feasible to simulate more than about 100 particles with current methods. Thus, development of reliable and more efficient simulation techniques, able to cope with thousands of suspended particles, would have a significant impact on our understanding of particulate suspensions, complementing present experimental and theoretical knowledge. In these two papers a new simulation technique for particulate suspensions is described; it combines Newtonian dynamics of the solid particles with a discretized Boltzmann model (McNamara and Zanetti, 1988; Higuera et al., 1989) for the fluid. The basic idea is illustrated in Fig. 1, which shows 5 solid particles suspended in a background fluid. The fluid can be modeled as a continuum (Fig. 1a), as a molecular liquid (Fig. 1b), or as a discrete velocity (lattice) gas (Fig. 1c). Because of the large scale separations, the dynamics of the solid particles are largely independent of the detailed mechanics of the suspending fluid. Discrete velocity models of the fluid combine most of the features of a fully molecular simulation of solid and liquid phases, but are orders of magnitude faster computationally. They have many advantages over conventional methods of simulating particulate suspensions, which are usually based on complicated, computationally intensive solutions of the Stokes equations. By contrast, lattice-gas/lattice-Boltzmann simulations are fast, flexible, and simple. The new method is closely related to earlier suspension modeling using lattice-gas cellular automata (Ladd et al., 1988; Ladd and Frenkel, 1989; Ladd and Frenkel, 1990; Ladd, 1991; van der Hoef et al., 1991), but the large and uncontrollable statistical fluctu-ations present in lattice-gas models are suppressed, reducing the need for computationally expensive ensemble averaging.

In the classical theory of suspensions (Happel and Brenner, 1986), the hydrodynamic interactions are assumed to be fully developed; in other words, there is a complete separation of time scales between the dynamics of the fluid and the motion of the solid particles. Most simulation methods (for instance Ermak and McCammon, 1978; Brady and Bossis, 1988; Ladd, 1988; Tran-Cong and Phan-Thien, 1989; Karrila et al., 1989) utilize this approximation, even though it imposes a crippling numerical burden, associated with the *global* nature of the interactions; in such cases one must either make further, drastic simplifications, as in Ermak and McCammon, 1978, or pay the steep computational cost of an algorithm that scales as at least the *square* and often as the *cube* of the number of particles (Brady and Bossis, 1988; Ladd, 1988; Tran-Cong and Phan-Thien, 1989; Karrila et al., 1989).[1,2] In reality, hydrodynamic interactions develop in time and space from purely *local* forces, generated at the solid-fluid surfaces, which then diffuse throughout the fluid. Recently a new simula-tion technique for particulate suspensions has been developed (Ladd, 1993b) which exploits this spatial and temporal locality; as a result the computational cost scales *linearly* with the number of particles. The method is also very flexible; the particle size and shape, the electrostatic interactions, the flow geometry, the Péclet number (the ratio of viscous forces to Brownian forces), and the Reynolds number (the ratio of inertial forces to viscous forces), can all be varied independently.

---

[1] The exact scaling depends on the problem to be solved. In some instances, particle velocities are computed for a given set of forces, in which case the computation can scale as $N^2$. However in many cases, for instance to simulate Brownian motion (Bossis and Brady, 1987), the full $6N \times 6N$ diffusion coefficient matrix is needed; here the computational cost is of order $N^3$. Moreover, determining lubrication forces (Durlofsky et al., 1987) also involves an order $N^3$ calculation of the friction coefficient matrix.

[2] The Stokes equations can be solved very efficiently by spectral methods (Fogelson and Peskin, 1988; Sulsky and Brackbill, 1991). However, there are then the usual difficulties associated with incorporating solid boundary conditions into spectral codes; in the referenced work the particle surfaces are represented by point forces. Numerical studies of the Navier-Stokes equations indicate that, even for pure fluid flows, the lattice-Boltzmann equation is quite competitive with the best spectral methods (Chen et al., 1992).

There is a well established connection between the dynamics of a dilute gas and the Navier-Stokes equations (Chapman and Cowling, 1960). Thus, if we can determine the time evolution of the one-particle velocity distribution function $n(\mathbf{r}, \mathbf{v}, t)$, which defines the density of particles with velocity $\mathbf{v}$ around the space-time point $(\mathbf{r}, t)$, then the hydrodynamic fields can be readily evaluated. By introducing the assumption of molecular chaos, *i.e.* that successive binary collisions in a dilute gas are uncorrelated, Boltzmann was able to derive the integro-differential equation for $n$ named after him (see Chapman and Cowling, 1960). The Navier-Stokes equations follow directly from the Boltzmann equation in the limit that the dimensions of the macroscopic flow fields are much larger than the mean free path between molecular collisions (Chapman and Cowling, 1960). Because of its complexity, there are few direct numerical solutions of the Boltzmann equation (an exception is Yen, 1984), but stochastic, particle-based simulations are quite commonly used in molecular gas dynamics; the merits of this approach are discussed in Bird, 1976 and Bird, 1990. A variant of this approach has recently been introduced to simulate particle suspensions (Hoogerbrugge and Koelman, 1992; Koelman and Hoogerbrugge, 1993). However, the one-particle distribution function $n(\mathbf{r}, \mathbf{v}, t)$ contains much more information than is strictly necessary to solve fluid-dynamics problems; hence, in recent years, there has been a resurgence of interest in discrete-velocity or lattice-gas models (Frisch et al., 1986; Frisch et al., 1987), in which the continuous distribution of molecular velocities is replaced by a few, carefully chosen, discrete values. An important consequence of this work is the realization that a Boltzmann equation, fully discretized in coordinate space, velocity space, and time, is equivalent to an explicit, second-order, finite-difference approximation to the Navier-Stokes equations (McNamara and Alder, 1992). Numerical studies have shown that a lattice-Boltzmann approximation can be comparable in accuracy and computational cost to state-of-the-art Navier-Stokes solvers, either finite difference (McNamara and Alder, 1992) or spectral (Chen et al., 1992).

In the lattice-Boltzmann approximation, the fundamental quantity is the discretized one-particle velocity distribution function $n_i(\mathbf{r}, t)$, which describes the number of particles at a particular node of the lattice $\mathbf{r}$, at a time $t$, with a velocity $\mathbf{c}_i$; $\mathbf{r}$, $t$, and $\mathbf{c}_i$ are discrete, whereas $n_i$ is continuous. The hydrodynamic fields, mass density $\rho$, momentum density $\mathbf{j} = \rho \mathbf{u}$, and momentum flux $\mathbf{\Pi}$, are moments of this velocity distribution:

$$\rho = \sum_i n_i, \quad \mathbf{j} = \sum_i n_i \mathbf{c}_i, \quad \mathbf{\Pi} = \sum_i n_i \mathbf{c}_i \mathbf{c}_i. \tag{1.1}$$

For simulations of particulate suspensions, the lattice-Boltzmann approximation has two important advantages over a finite-difference approximation to the Navier-Stokes equations. First, the connection to molecular mechanics makes it possible to derive simple local rules for the interactions between the fluid and the suspended solid particles (Ladd, 1993b); this was demonstrated in our earlier lattice-gas simulations (Ladd and Frenkel, 1990; Ladd, 1991; van der Hoef et al., 1991). Second, the discrete one-particle distribution function $n_i$ contains additional information about the dynamics of the fluid beyond that contained in the Navier-Stokes equations; in particular, the fluid stress tensor, although dynamically coupled to the velocity gradient (Frisch et al., 1987), has an independent significance at short times. This additional flexibility allows us to simulate molecular fluctuations, leading to Brownian motion of the suspended particles. To do this, a random fluctuation, uncorrelated in space and time, is added to the fluid stress tensor (Ladd, 1993b); the variance of the fluctuations defines the effective temperature of the fluctuating fluid (Landau and Lifshitz, 1959). This approach is quite different from Brownian dynamics (Ermak and McCammon, 1978) or Stokesian dynamics (Bossis and Brady, 1987), where random fluctuations are applied directly to the particles; to include hydrodynamic interactions in these methods requires sampling from the $6N \times 6N$ diffusion matrix, which is extremely time consuming.

The layout of the remainder of this paper is as follows. In section II the lattice-Boltzmann approximation is described and its connection to Navier-Stokes fluid dynamics is established. Section III describes the implementation of the solid-fluid boundary conditions at the microscopic level, together with analytic and numerical results for shear flows and channel flows. In section IV fluctuations are introduced into the lattice-Boltzmann approximation; it is verified that the simulations satisfy the fluctuation-dissipation theorem and that the correct shear viscosity can be obtained from an appropriate Green-Kubo formulae. In the companion paper (Ladd, 1993a), referred to hereafter as paper II, results of extensive tests of creeping-flow hydrodynamics are reported, for both periodic arrays and random distributions of spheres; time-dependent and finite Reynolds number flows are also discussed. Three-dimensional simulations of up to 1024 colloidal particles, moving under the action of Brownian forces, are also reported.

## II. DISCRETE BOLTZMANN APPROXIMATION

The computational utility of the lattice-Boltzmann equation is related to the realization that only a small set of discrete velocities are necessary to simulate the Navier-Stokes equations (Frisch et al., 1986). It may be helpful in what follows to imagine an underlying mechanical model in which identical particles move with discrete velocities from node to node of a regular lattice; much of the kinetic theory dilute gases can then be carried over directly to the discretized version. The specific model used in this work has 18 different velocities corresponding to the near-neighbor and second-neighbor directions of a simple cubic lattice. Thus there are six velocities of speed 1, corresponding to (100) directions in the lattice and 12 velocities of speed $\sqrt{2}$, corresponding to the (110) directions, for a total of 18. All quantities in this paper are expressed in "lattice units", for which the distance between nearest-neighbor nodes and the time for the particles to travel from node to node are both unity. Note that the velocities are such that all particles move from node to node simultaneously.

The time evolution of the distribution functions $n_i$ is described by a discrete analogue of the Boltzmann equation (Frisch et al., 1987),

$$n_i(\mathbf{r} + \mathbf{c}_i, t + 1) = n_i(\mathbf{r}, t) + \Delta_i(\mathbf{r}, t), \qquad (2.1)$$

where $\Delta_i$ is the change in $n_i$ due to instantaneous molecular collisions at the lattice nodes. The post collision distribution $n_i + \Delta_i$ is propagated for one time step, in the direction $\mathbf{c}_i$. The collision operator $\Delta(n)$ depends on all the $n_i$'s at the node, denoted collectively by $n(\mathbf{r}; t)$; it can take any form, subject to the constraints of mass and momentum conservation. An exact expression for the Boltzmann collision operator has been derived for several different lattice-gas models (Frisch et al., 1987; McNamara and Zanetti, 1988), under the usual assumption that the distribution functions $n(\mathbf{r}; t)$ are uncorrelated from those at previous times. However, such collision operators are complex and ill-suited to numerical simulation. A computationally useful form for the collision operator can be constructed by linearizing the collision operator about the local equilibrium $n^{eq}$ (Higuera et al., 1989), i.e.

$$\Delta_i(n) = \Delta_i(n^{eq}) + \sum_j \mathcal{L}_{ij}(n_j - n_j^{eq}), \qquad (2.2)$$

where $\mathcal{L}$ is the linearized collision operator, and $\Delta_i(n^{eq}) = 0$ by definition. It is not necessary to construct a particular collision operator and from this calculate $\mathcal{L}$; rather it is sufficient to consider the general principles of conservation and symmetry and then to construct the eigenvalues and eigenvectors of $\mathcal{L}$. However, before doing this, we will determine the proper form for the equilibrium distribution function.

To establish the connection between molecular mechanics and fluid dynamics, it is usual to split the distribution function into an equilibrium part and a non-equilibrium part

$$n_i = n_i^{eq} + n_i^{neq}. \qquad (2.3)$$

The equilibrium distribution is a collisional invariant (i.e. $\Delta_i(n^{eq}) = 0$), and depends only on the local hydrodynamic variables (mass density, stream velocity and, in some cases, energy density); for a molecular gas $n^{eq}$ is the Maxwell-Boltzmann distribution. It is well known that a Maxwell-Boltzmann local equilibrium leads to the Euler equations of hydrodynamics (Chapman and Cowling, 1960); however the most-probable (equilibrium) distribution functions of discrete velocity lattice gases give rise to density dependent advection velocities and velocity dependent pressures (Frisch et al., 1987). Therefore we seek a constrained equilibrium distribution for our discretized velocity model, that will lead to the correct macroscopic fluid dynamics at the Euler level; the required form for the moments of the velocity distribution function defined in Eq. (1.1) are

$$\rho = \sum_i n_i^{eq}, \quad \mathbf{j} = \sum_i n_i^{eq} \mathbf{c}_i, \quad \mathbf{\Pi}^{eq} = \sum_i n_i^{eq} \mathbf{c}_i \mathbf{c}_i = p\mathbf{1} + \rho \mathbf{u}\mathbf{u}, \tag{2.4}$$

where $p$ is the pressure and $\mathbf{\Pi}^{eq}$ is the non-dissipative part of the momentum flux. As in the usual kinetic theory of gases, the viscous fluxes come from the non-equilibrium part of the distribution function.

The equilibrium distribution can be expressed as a series expansion in powers of the flow velocity $\mathbf{u}$,

$$n_i^{eq} = \rho \left[ a_0^{c_i} + a_1^{c_i} \mathbf{u} \cdot \mathbf{c}_i + a_2^{c_i} \overline{\mathbf{u}\mathbf{u}} : \overline{\mathbf{c}_i \mathbf{c}_i} + a_3^{c_i} u^2 \right], \tag{2.5}$$

where $\overline{\mathbf{u}\mathbf{u}} = \mathbf{u}\mathbf{u} - (1/3)u^2 \mathbf{1}$ is the traceless part of $\mathbf{u}\mathbf{u}$; Eq. 2.5 has the same functional form as a small $u$ expansion of the Maxwell-Boltzmann distribution function. The moments of the distribution function (Eq. 2.4) can be expressed in terms of the coefficients $a_i^{c_i}$, which are functions only of the speed, $c_i$;

$$\rho^{-1} \sum_i n_i^{eq} \quad = 6a_0^1 + 12a_0^{\sqrt{2}} + \left[ 6a_3^1 + 12a_3^{\sqrt{2}} \right] u^2, \tag{2.6}$$

$$\rho^{-1} \sum_i n_i^{eq} c_{i\alpha} \quad = [2a_1^1 + 8a_1^{\sqrt{2}}] u_\alpha, \tag{2.7}$$

$$\rho^{-1} \sum_i n_i^{eq} c_i^2 \quad = 6a_0^1 + 24a_0^{\sqrt{2}} + \left[ 6a_3^1 + 24a_3^{\sqrt{2}} \right] u^2, \tag{2.8}$$

$$\rho^{-1} \sum_i n_i^{eq} \overline{c_{i\alpha} c_{i\beta}} = \left\{ 2a_2^1 [\delta_{\alpha\beta\gamma\delta} - (1/3)\delta_{\alpha\beta}\delta_{\gamma\delta}] \right.$$
$$\left. + 4a_2^{\sqrt{2}} [-\delta_{\alpha\beta\gamma\delta} - (1/3)\delta_{\alpha\beta}\delta_{\gamma\delta} + \delta_{\alpha\gamma}\delta_{\beta\delta} + \delta_{\alpha\delta}\delta_{\beta\gamma}] \right\} \overline{u_\gamma u_\delta}, \tag{2.9}$$

$$\rho^{-1} \sum_i n_i^{eq} c_{i\alpha} c_{i\beta} c_{i\gamma} = \left\{ 2a_1^1 \delta_{\alpha\beta\gamma\delta} \right.$$
$$\left. + 4a_1^{\sqrt{2}} [-\delta_{\alpha\beta\gamma\delta} + \delta_{\alpha\beta}\delta_{\gamma\delta} + \delta_{\alpha\gamma}\delta_{\beta\delta} + \delta_{\alpha\delta}\delta_{\beta\gamma}] \right\} u_\delta. \tag{2.10}$$

The tensor $\delta_{\alpha\beta\gamma\delta}$ is unity when all the subscripts are the same (*i.e.* $\delta_{xxxx} = \delta_{yyyy} = \delta_{zzzz} = 1$) and zero otherwise; $\delta_{\alpha\beta}$ is the Kronecker delta. The third-order moments (Eq. 2.10) do not contribute to the Euler equations, but they must be proportional to the fourth-rank identity tensor $(\delta_{\alpha\beta}\delta_{\gamma\delta} + \delta_{\alpha\gamma}\delta_{\beta\delta} + \delta_{\alpha\delta}\delta_{\beta\gamma})$; otherwise the viscous part of the momentum flux will not be isotropic (see Eq. 2.28). A comparison of Eqs. 2.6–2.9 with Eq. 2.4, together with the isotropy condition Eq. 2.10, is sufficient to determine all the coefficients,

$$\begin{aligned} a_0^1 &= (2 - 3c_s^2)/6, & a_1^1 &= 1/6, & a_2^1 &= 1/4, & a_3^1 &= -1/6, \\ a_0^{\sqrt{2}} &= (3c_s^2 - 1)/12, & a_1^{\sqrt{2}} &= 1/12, & a_2^{\sqrt{2}} &= 1/8, & a_3^{\sqrt{2}} &= 1/12. \end{aligned} \quad (2.11)$$

The definition of the pressure (Eq. 2.4) indicates that it is proportional to the density, *i.e.* $p = (1/3)(\sum_i n_i^{eq} c_i^2 - \rho u^2) = \rho c_s^2$; later it will be shown that $c_s$ is the speed of sound (see Eq. 2.26), as in a normal gas. Since the distribution function must always be positive, the speed of sound is bounded by the limits $1/3 \leq c_s^2 \leq 2/3$. However, as the sound speed approaches either of the two limits, the simulation becomes unstable with respect to variations in fluid velocity. In our simulations, the intermediate value $c_s = \sqrt{1/2}$ is used, to maximize the stability with respect to variations in flow velocity; in this case

$$a_0^1 = 1/12, \quad a_0^{\sqrt{2}} = 1/24. \qquad (2.12)$$

Having constructed an equilibrium distribution appropriate for the inviscid (Euler) equations, let us consider next how to obtain the correct form for the viscous terms in the fluid equations. We require that the linearized collision operator satisfy the following eigenvalue equations;

$$\sum_i \mathcal{L}_{ij} = 0, \quad \sum_i \mathbf{c}_i \mathcal{L}_{ij} = 0, \quad \sum_i \overline{\mathbf{c}_i \mathbf{c}_i} \mathcal{L}_{ij} = \lambda \overline{\mathbf{c}_j \mathbf{c}_j}, \quad \sum_i c_i^2 \mathcal{L}_{ij} = \lambda_B c_j^2. \qquad (2.13)$$

The first two equations follow from conservation of mass and momentum and the last two equations describes the isotropic relaxation of the stress tensor; the eigenvalues $\lambda$ and $\lambda_B$ are related to the shear and bulk viscosities (Eq. 2.34). Equations 2.13 account for 10 of the 18 eigenvectors of $\mathcal{L}$. The remaining 8 modes, comprising some higher-order moments of $\mathcal{L}$, are not relevant to simulations of the Navier-Stokes equations and will be ignored. Their eigenvalues are set to -1 so that these modes are then projected out entirely from the post-collision distribution; this both simplifies the simulation and ensures the fastest possible relaxation of the non-hydrodynamic modes. The computational procedure to update the lattice-Boltzmann equation is therefore quite straightforward. At each site the moments $\rho$, $\mathbf{j}$, and $\mathbf{\Pi}$ (Eq. 1.1), and the equilibrium momentum flux $\mathbf{\Pi}^{eq}$ (Eq. 2.4) are calculated. The momentum flux is updated according to Eq. 2.13

$$\Pi'_{\alpha\beta} = \Pi^{eq}_{\alpha\beta} + (1+\lambda)(\overline{\Pi}_{\alpha\beta} - \overline{\Pi}^{eq}_{\alpha\beta}) + (1+\lambda_B)(\Pi_{\gamma\gamma} - \Pi^{eq}_{\gamma\gamma})(1/3)\delta_{\alpha\beta}; \quad (2.14)$$

then the post-collision distribution, $n_i + \Delta_i(n)$, is determined from the inverse of Eq. 1.1,

$$n_i + \Delta_i(n) = a_0^{c_i}\rho + a_1^{c_i} j_\alpha c_{i\alpha} + a_2^{c_i} \Pi'_{\alpha\beta}\overline{c_{i\alpha}c_{i\beta}} + a_3^{c_i}(\Pi'_{\alpha\alpha} - 3\rho c_s^2). \quad (2.15)$$

The term $-3\rho c_s^2 a_3^{c_i}$ keeps $\mathbf{\Pi}$ orthogonal to $\rho$.

Next we examine the macrodynamical behavior arising from the lattice-Boltzmann equation; our method of solution is the usual multi-time-scale analysis (Frisch et al., 1987). We begin with conservation equations for the moments of the distribution function

$$\sum_i n_i(\mathbf{r}+\mathbf{c}_i, t+1) = \sum_i n_i(\mathbf{r}, t), \quad (2.16)$$

$$\sum_i n_i(\mathbf{r}+\mathbf{c}_i, t+1)c_{i\alpha} = \sum_i n_i(\mathbf{r}, t)c_{i\alpha}, \quad (2.17)$$

$$\sum_i n_i(\mathbf{r}+\mathbf{c}_i, t+1)\overline{c_{i\alpha}c_{i\beta}} = \sum_i n_i(\mathbf{r}, t)\overline{c_{i\alpha}c_{i\beta}} + \lambda\sum_i n_i^{neq}(\mathbf{r}, t)\overline{c_{i\alpha}c_{i\beta}}. \quad (2.18)$$

$$\sum_i n_i(\mathbf{r}+\mathbf{c}_i, t+1)c_i^2 = \sum_i n_i(\mathbf{r}, t)c_i^2 + \lambda_B\sum_i n_i^{neq}(\mathbf{r}, t)c_i^2. \quad (2.19)$$

To find the long-time, long-wavelength dynamics, we introduce a scaling parameter $\epsilon$, defined as the ratio of the lattice spacing to a characteristic macroscopic length; the hydrodynamic limit corresponds to $\epsilon \ll 1$. In a molecular gas the appropriate scaling parameter is the Knudsen number, the ratio of the mean-free path between collisions to the macroscopic length scale. The parameter $\epsilon$ plays a similar role to the Knudsen number in the Chapman-Enskog method (Chapman and Cowling, 1960); it is used, first of all, to separate the relaxation of the equilibrium and non-equilibrium distributions, (*c.f.* Eq. 2.3)

$$n_i = n_i^{eq} + \epsilon n_i^{neq}. \quad (2.20)$$

However, because the lattice spacing and the mean-free path are comparable, there are additional contributions to the viscous momentum flux, which do not appear in the ordinary kinetic theory of gases (see Eq. 2.32). In order to remove discrete lattice artifacts from the macroscopic equations, it is convenient to define a macroscopic space scale $\mathbf{r}_1 = \epsilon \mathbf{r}$, and two macroscopic time scales $t_1 = \epsilon t$ and $t_2 = \epsilon^2 t$; this enables a separation to be made between the propagation of sound ($t_1$) and the diffusion of vorticity ($t_2$) (Frisch et al., 1987).

Expanding the finite differences, $n_i(\mathbf{r} + \mathbf{c}_i, t+1) - n_i(\mathbf{r}, t)$, in Eqs. 2.16–2.19 about $\mathbf{r}$ and $t$, and collecting terms that are first order in $\epsilon$ we obtain the relaxation on the $t_1$ time scale;

$$\partial_{t_1} \sum_i n_i^{eq} + \nabla_\alpha \sum_i n_i^{eq} c_{i\alpha} = 0, \tag{2.21}$$

$$\partial_{t_1} \sum_i n_i^{eq} c_{i\alpha} + \nabla_\beta \sum_i n_i^{eq} c_{i\alpha} c_{i\beta} = 0, \tag{2.22}$$

$$\partial_{t_1} \sum_i n_i^{eq} c_{i\alpha} c_{i\beta} + \nabla_\gamma \sum_i n_i^{eq} c_{i\alpha} c_{i\beta} c_{i\gamma} = \lambda \sum_i n_i^{neq} \overline{c_{i\alpha} c_{i\beta}} + \lambda_B \sum_i n_i^{neq} c_i^2/3; \tag{2.23}$$

the gradient operator refers to derivatives on the macroscopic $\mathbf{r}_1$ space scale, it i.e. $\boldsymbol{\nabla} \equiv \partial_{\mathbf{r}_1}$.

The equations for mass and momentum conservation (Eqs. 2.21 and 2.22) can be rewritten using Eq. 2.4;

$$\partial_{t_1} \rho + \boldsymbol{\nabla} \cdot (\rho \mathbf{u}) = 0, \tag{2.24}$$

$$\partial_{t_1}(\rho \mathbf{u}) + \boldsymbol{\nabla} \cdot (p\mathbf{1} + \rho \mathbf{u}\mathbf{u}) = 0, \tag{2.25}$$

which are the Euler equations of hydrodynamics. Substituting the equation of state $p = \rho c_s^2$ and linearizing the Euler equations with respect to $\delta\rho$ and $\mathbf{u}$, we find that, for small density fluctuations,

$$\partial_{t_1}^2 \rho = c_s^2 \nabla^2 \rho. \tag{2.26}$$

Equation 2.26 shows that density fluctuations relax via the propagation of sound waves, on a time scale $t_1$, and therefore decouple from the $t_2$ time scale evolution of the viscous stresses. The time derivative that appears in Eq. 2.23 can be evaluated by using Eqs. 2.24 and 2.25 to express the time derivatives of $\rho$ and $\rho \mathbf{u}$ in terms of spatial derivatives;

$$\partial_{t_1} \sum_i n_i^{eq} c_{i\alpha} c_{i\beta} = \partial_{t_1}(\rho c_s^2 \delta_{\alpha\beta} + \rho u_\alpha u_\beta)$$
$$= -\nabla_\gamma(\rho u_\alpha u_\beta u_\gamma) - c_s^2 [u_\alpha \nabla_\beta \rho + u_\beta \nabla_\alpha \rho + \nabla_\gamma(\rho u_\gamma)\delta_{\alpha\beta}]. \quad (2.27)$$

The spatial derivative of the third-order moment can be evaluated directly from Eqs. 2.10 and 2.11,

$$\nabla_\gamma \sum_i n^{eq} c_{i\alpha} c_{i\beta} c_{i\gamma} = (1/3)[\nabla_\alpha(\rho u_\beta) + \nabla_\beta(\rho u_\alpha) + \nabla_\gamma(\rho u_\gamma)\delta_{\alpha\beta}]. \quad (2.28)$$

In the incompressible limit, variations in density can be ignored, so that

$$\partial_{t_1} \sum_i n_i^{eq} \overline{c_{i\alpha} c_{i\beta}} + \nabla_\gamma \sum_i n_i^{eq} \overline{c_{i\alpha} c_{i\beta}} c_{i\gamma} = (\rho/3)[\nabla_\alpha u_\beta + \nabla_\beta u_\alpha - (2/3)\nabla_\gamma u_\gamma], \quad (2.29)$$

with errors of order $\nabla u^3$. Then, from Eq. 2.23, the Navier-Stokes form for the viscous stresses can be obtained,

$$\sigma_{\alpha\beta} = -\sum_i n_i^{neq} \overline{c_{i\alpha} c_{i\beta}} = -(\rho/3\lambda)(\nabla_\alpha u_\beta + \nabla_\beta u_\alpha - (2/3)\nabla_\gamma u_\gamma \delta_{\alpha\beta}); \quad (2.30)$$

hereafter the non-equilibrium (viscous) contributions to the pressure will be ignored. Since the Mach number in our simulations is typically $10^{-2}$ or less, we can safely ignore the effects of compressibility. It is interesting to note that if the speed of sound $c_s$ were set to $\sqrt{1/3}$ instead of $\sqrt{1/2}$, then inspection of Eqs. 2.27 and 2.28 indicates that the correct form for the viscous stresses, including the non-equilibrium pressure, would be obtained (with corrections of order $\nabla u^3$), even for non-zero Mach numbers. For such simulations, an additional density of zero-velocity particles is required to maintain stability (McNamara and Alder, 1993).

The $t_2$ relaxation of the mass and momentum densities can be found from the order $\epsilon^2$ terms in the expansion of Eqs. 2.16 and 2.17;

$$\partial_{t_2} \sum_i n_i^{eq} = \partial_{t_2}\rho = 0, \quad (2.31)$$

$$\partial_{t_2} \sum_i n_i^{eq} c_{i\alpha} + (1/2)\nabla_\beta(\partial_{t_1} \sum_i n_i^{eq} c_{i\alpha} c_{i\beta} + \nabla_\gamma \sum_i n_i^{eq} c_{i\alpha} c_{i\beta} c_{i\gamma}) + \nabla_\beta \sum_i n_i^{neq} c_{i\alpha} c_{i\beta} = 0. \quad (2.32)$$

Equation 2.31 shows that the fluid is incompressible on the $t_2$ time scale; all relaxation of density fluctuations takes place on the $t_1$ time scale. Using Eqs. 2.29 and 2.30 we can express the long-time variation of the viscous stresses in a form identical to the incompressible Navier-Stokes equations,

$$\partial_{t_2}(\rho \mathbf{u}) = \eta \nabla^2 \mathbf{u} \tag{2.33}$$

where

$$\eta = -\rho(2/\lambda + 1)/6 \tag{2.34}$$

is the shear viscosity; onece again terms proportional to $\boldsymbol{\nabla} \cdot \mathbf{u}$ are neglected.

The shear viscosity (Eq. 2.34) contains two distinct contributions: the first, proportional to $\lambda^{-1}$, arises from molecular-like collisions (Eq. 2.13); the second term comes from the diffusion of momentum caused by the finite lattice (Eq. 2.29). In most situations of practical interest, the collisional and lattice contributions to the viscosity are of comparable magnitude; fortunately they both have exactly the same dependence on velocity gradient, so that they may be combined into a single transport coefficient. A linear stability analysis shows that $\lambda$ must be bounded in the range $-2 < \lambda < 0$ (Higuera et al., 1989; McNamara and Alder, 1993), otherwise the shear stress grows exponentially in time. The bounds on $\lambda$ correspond to the simple physical requirement that the viscosity is positive. Actually the viscosity is bounded by a more stringent non-linear stability criterion, which has not yet been worked out in detail. Essentially the quantity $v^2/\eta$, where $v$ is a characteristic flow velocity, must be larger than some positive constant; this means there must always be some small but finite damping of the viscous modes.

Combining the relaxation of the momentum density on the $t_1$ and $t_2$ time scales, we obtain the incompressible Navier-Stokes equation

$$\partial_t(\rho \mathbf{u}) + \boldsymbol{\nabla} \cdot (\rho \mathbf{u} \mathbf{u}) = -\boldsymbol{\nabla} p + \eta \nabla^2 \mathbf{u}, \tag{2.35}$$

with equation of state $p = \rho c_s^2$. Once again we again point out that for the 18 velocity model used in this work, the simulations are only valid at low Mach numbers; slightly more complex models are needed to capture compressibility effects correctly. In the remainder of this paper, it will be assumed that the simulations will be run under conditions of low Mach number, with particle velocities $\mathbf{U}$ much less than the sound speed $c_s$; thus $\boldsymbol{\nabla} \cdot \mathbf{u} = 0$ to a good approximation.

Many flows involving particulate suspensions occur at low Reynolds number, and can be modeled by the creeping-flow or Stokes equations

$$\boldsymbol{\nabla} \cdot \mathbf{u} = 0, \quad \boldsymbol{\nabla} p = \eta \nabla^2 \mathbf{u}; \tag{2.36}$$

or, in terms of the momentum density $\mathbf{j} = \rho \mathbf{u}$,

$$\boldsymbol{\nabla} \cdot \mathbf{j} = 0, \quad \boldsymbol{\nabla} p = \nu \nabla^2 \mathbf{j}. \tag{2.37}$$

In our simulations, we do not model the Stokes equations directly, but rather as a limit (for small particle velocities) of the linearized, incompressible Navier-Stokes equation

$$\partial_t \mathbf{j} = -\boldsymbol{\nabla} p + \nu \nabla^2 \mathbf{j}; \tag{2.38}$$

Eq. 2.38 can be simulated directly by a change in the equilibrium distribution (*c.f.* Eq. 2.5),

$$n_i^{eq} = a_0^{c_i} \rho + a_1^{c_i} \mathbf{j} \cdot \mathbf{c}_i. \tag{2.39}$$

Finally, a significant simplification of the code occurs when $\lambda = -1$, corresponding to a viscosity $\eta = \rho/6$. Although such a large viscosity is not suitable for high Reynolds number flows, in the creeping flow limit it allows for a considerable simplification of the collision operator

$$n_i + \Delta_i(n) = a_0^{c_i} \rho + a_1^{c_i} \mathbf{j} \cdot \mathbf{c}_i, \tag{2.40}$$

which requires about half the number of floating-point operations as Eq. 2.15; most of our $R_e = 0$ simulations use this viscosity.

## III. SOLID-FLUID BOUNDARY CONDITIONS

To simulate the hydrodynamic interactions between solid particles in suspension, the lattice-Boltzmann approximation must be modified to incorporate the boundary conditions imposed on the fluid by the solid particles. The basic methodology is illustrated in Fig. 2. The solid particles are defined by a boundary surface, which can be of any size or shape; in Fig. 2 it is a circle. When placed on the lattice, the boundary surface cuts some of the links between lattice nodes. The fluid particles moving along these links interact with the solid surface at boundary nodes placed halfway along the links. Thus we obtain a discrete representation of the particle surface, which becomes more and more precise as the particle gets larger. Lattice nodes on either side of the boundary surface are treated in an identical fashion, so that fluid fills the whole volume of space, both inside and outside the solid particles. However, because of the relatively small volume inside each particle, the interior fluid relaxes quite quickly to rigid-body motion, characterized by the particle velocity and angular velocity. Thus, on physically important time scales, the interior fluid only contributes an added inertial mass to the solid particle.

In comparison with our previous work (Ladd et al., 1988; Ladd and Frenkel, 1989; Ladd and Frenkel, 1990; Ladd, 1991; van der Hoef et al., 1991), here we have chosen to place the boundary nodes on the links connecting the interior and exterior regions, whereas in our lattice-gas simulations they were located on the nodes closest to the boundary surface. There is little to choose between the two methods; the link method has the advantage that it provides a somewhat higher resolution of the solid boundary surface, as can be seen (Fig. 2) from the much larger number of boundary nodes compared with the number of lattice nodes just inside the surface. On the other hand the node method is faster, although this is of little significance in the computationally more intensive lattice-Boltzmann (as opposed to lattice-gas) simulations. Although at this point in time, our lattice-Boltzmann simulations have been limited to simple symmetrical objects; spheres, disks and plane walls, this restriction is not fundamental: in fact a limited number of lattice-gas simulations containing elongated objects have already been reported (van der Hoef, 1992).

At each boundary node there are two incoming distributions $n_i(\mathbf{r}, t_+)$ and $n_{i'}(\mathbf{r} + \mathbf{c}_i, t_+)$, corresponding to velocities $\mathbf{c}_i$ and $\mathbf{c}_{i'}$ ($\mathbf{c}_{i'} = -\mathbf{c}_i$) parallel to the link connecting $\mathbf{r}$ and $\mathbf{r} + \mathbf{c}_i$; the notation $n_i(\mathbf{r}, t_+) = n_i(\mathbf{r}, t) + \Delta_i(\mathbf{r}, t)$ is used to indicate the post-collision distribution (Eq. 2.15). In some cases boundary nodes for two different link directions, perpendicular to one another, may be coincident (see Fig. 2); these are treated independently. The velocity of the boundary node $\mathbf{u}_b$ is determined by the solid particle velocity $\mathbf{U}$, angular velocity $\mathbf{\Omega}$, and centre of mass $\mathbf{R}$,

$$\mathbf{u}_b = \mathbf{U} + \mathbf{\Omega} \times (\mathbf{r} + \tfrac{1}{2}\mathbf{c}_i - \mathbf{R}). \tag{3.1}$$

By exchanging population density between $n_i$ and $n_{i'}$ we can modify the local momentum density of the fluid to match the velocity of the solid particle surface at the boundary node, without affecting either the mass density or the stress, which depend only on the sum $n_i + n_{i'}$. Because the stress tensor is unaffected by the boundary-node interactions, it then follows that the hydrodynamic stick boundary condition applies right up to the solid surface, without any intervening boundary layer (Ladd and Frenkel, 1990). This point will be illustrated in more detail later. The mechanism for the boundary-node interactions is illustrated in Fig. 3. In Fig. 3a we see the two incoming populations, $n_i(\mathbf{r}, t_+)$ and $n_{i'}(\mathbf{r} + \mathbf{c}_i, t_+)$, interacting with a stationary boundary node. In this case, the populations are simply reflected back in the direction they came from (Frisch et al., 1987; Cornubert et al., 1991), so that

$$n_i(\mathbf{r} + \mathbf{c}_i, t+1) = n_{i'}(\mathbf{r} + \mathbf{c}_i, t_+), \quad \text{and} \quad n_{i'}(\mathbf{r}, t+1) = n_i(\mathbf{r}, t_+). \tag{3.2}$$

In Figs 3b and 3c the effects of a moving object can be seen. In addition to reflection, population density is now transferred across the boundary node, in proportion to the velocity of the node $\mathbf{u}_b$,

$$n_i(\mathbf{r} + \mathbf{c}_i, t+1) = n_{i'}(\mathbf{r} + \mathbf{c}_i, t_+) + 2a_1^{c_i}\rho \mathbf{u}_b \cdot \mathbf{c}_i,$$
$$n_{i'}(\mathbf{r}, t+1) = n_i(\mathbf{r}, t_+) - 2a_1^{c_i}\rho \mathbf{u}_b \cdot \mathbf{c}_i; \tag{3.3}$$

these results are ensemble averages of our earlier boundary-node collision rules for lattice gases (Ladd and Frenkel, 1989; Ladd, 1991). Only the velocity component of the boundary node along the link direction ($\mathbf{c}_i$) is included in the calculation of population transfer; thus the outcome in Figs. 3b and 3c is the same. The general form for the boundary node interactions in Eq. 3.3 is determined by the requirement that the local mass density and stress tensor are conserved; thus rearrangements of population can only be made among pairs of opposite velocities. Furthermore, for stationary nodes we must recover the usual "bounce-back" condition (Eq. 3.2). The exact amount of population density transferred (*i.e* the magnitude of the $\mathbf{u}_b \cdot \mathbf{c}_i$ term) is determined by the requirement that any distribution consistent with the boundary-node velocity $\mathbf{u}_b$ is stationary with respect to interactions with the boundary nodes. It is not obvious that Eq. 3.3 satisfies this condition, but we will verify that this is indeed so in the next paragraph.

Let us now examine boundary-node interactions in more detail. From section II we know that the distribution function at a node can be written as the sum of equilibrium (Eq. 2.5) and non-equilibrium contributions (Eq. 2.30)

$$n_i^{neq} = -a_2^{c_i} \boldsymbol{\sigma} : \overline{\mathbf{c}_i \mathbf{c}_i}. \tag{3.4}$$

Ignoring higher order gradients (*i.e.* $\nabla \nabla u$) and terms proportional to $\boldsymbol{\nabla} \cdot \mathbf{u}$, the collisional stress tensor in Eq. 3.4 can be expressed in terms of velocity gradients (Eq. 2.30), so that

$$\lambda n_i^{neq} = a_1^{c_i} \rho \mathbf{c}_i \mathbf{c}_i : \boldsymbol{\nabla} \mathbf{u}. \tag{3.5}$$

Then the post-collision distribution $n_i(\mathbf{r}, t_+) = n_i^{eq}(\mathbf{r}, t) + (1 + \lambda) n_i^{neq}(\mathbf{r}, t)$ (*c.f.* Eqs. 2.14 and 2.15) is given, to the same approximation, by

$$n_i(\mathbf{r}, t_+) = n_i(\mathbf{r}, t) + a_1^{c_i} \rho \mathbf{c}_i \mathbf{c}_i : \boldsymbol{\nabla} \mathbf{u}(\mathbf{r}) = n_{i'}(\mathbf{r}, t) + 2 a_1^{c_i} \rho \mathbf{u}(\mathbf{r}) \cdot \mathbf{c}_i + a_1^{c_i} \rho \mathbf{c}_i \mathbf{c}_i : \boldsymbol{\nabla} \mathbf{u}(\mathbf{r}), \tag{3.6}$$

where we have exploited the symmetries in the distribution functions for velocities $i$ and $i'$. If there is a boundary node located at $\mathbf{r} + \frac{1}{2} \mathbf{c}_i$, then the population $n_{i'}(\mathbf{r}, t+1)$ is modified according to Eq. 3.3, *i.e.*

$$n_{i'}(\mathbf{r}, t+1) = n_{i'}(\mathbf{r}, t) + 2 a_1^{c_i} \rho \left[ \mathbf{u}(\mathbf{r}) + (1/2) \mathbf{c}_i \cdot \boldsymbol{\nabla} \mathbf{u}(\mathbf{r}) - \mathbf{u}_b(\mathbf{r} + \tfrac{1}{2} \mathbf{c}_i) \right] \cdot \mathbf{c}_i; \tag{3.7}$$

thus the distribution is stationary when the fluid velocity $\mathbf{u}(\mathbf{r} + \tfrac{1}{2} \mathbf{c}_i) = \mathbf{u}(\mathbf{r}) + (1/2) \mathbf{c}_i \cdot \boldsymbol{\nabla} \mathbf{u}(\mathbf{r})$ is equal to the boundary-node velocity $\mathbf{u}_b$.

To illustrate the action of the boundary nodes more clearly, we consider, as an explicit example, planar Couette flow. Figure 4 shows a two-dimensional projection of the lattice-Boltzmann model onto the $xy$-plane; the system is assumed to be time independent, and translationally invariant in the $y$- and $z$-directions. As an idealized model of a solid particle surface, two infinite planes of boundary nodes are set up, at $x = 0$ and $x = L$. In the fluid between the boundary surfaces ($0 < x < L$) there is a uniform velocity gradient $\nabla_x u_y(x) = \gamma$; outside the boundary planes, the fluid moves with uniform velocity equal to the wall velocity. Note that in this example, the lattice nodes are more conveniently set at half-integer values of $x$. The problem then, is to find the distribution function for this flow geometry that is stationary under the action of the boundary-node microrules, with velocities $\mathbf{u}_b(x = 0) = 0$ and $\mathbf{u}_b(x = L) = \gamma L$. A similar problem, involving mixed stick-slip boundary conditions at a stationary wall, has been addressed by Cornubert et al., 1991.

The expected velocity distribution in a uniform velocity gradient can be constructed from the equilibrium distribution for Stokes flow (Eq. 2.39) and the non-equilibrium distribution Eq. 3.5,

$$n_i^{neq} = \rho \gamma c_{ix} c_{iy} / 12. \tag{3.8}$$

Using the notation of Fig. 4, the velocity distribution function in the fluid ($0 < x < L$) can be written explicitly as

$$\begin{aligned}
n_0(x) &= 4, \\
n_1(x) &= 4, & n_2(x) &= 4, \\
n_3(x) &= 4(1 + 2\gamma x), & n_4(x) &= 4(1 - 2\gamma x), \\
n_5(x) &= (1 + 2\gamma x + 2\gamma/\lambda), & n_6(x) &= (1 - 2\gamma x + 2\gamma/\lambda), \\
n_7(x) &= (1 - 2\gamma x - 2\gamma/\lambda), & n_8(x) &= (1 + 2\gamma x - 2\gamma/\lambda);
\end{aligned} \tag{3.9}$$

the densities $n_0$ through $n_4$ have been multiplied by 4 to account for the number of projected velocities, and the mass density has, for convenience, been set equal to 24. The velocity distribution away from the boundaries can be updated according to the usual time evolution of the lattice-Boltzmann equation. From Eqs. 1.1, 2.14, and 2.15, we can compute the post-collision distribution and then propagate it to the neighboring nodes using Eq. 2.1. The new velocity distribution, denoted by $n'$, is

$$\begin{aligned}
n'_0(x) &= 4, \\
n'_1(x) &= 4, & n'_2(x) &= 4, \\
n'_3(x) &= 4(1 + 2\gamma x), & n'_4(x) &= 4(1 - 2\gamma x), \\
n'_5(x) &= [1 + 2\gamma(x - 1) + 2\gamma(1 + \lambda)/\lambda], & n'_6(x) &= [1 - 2\gamma(x + 1) + 2\gamma(1 + \lambda)/\lambda], \\
n'_7(x) &= [1 - 2\gamma(x - 1) - 2\gamma(1 + \lambda)/\lambda], & n'_8(x) &= [1 + 2\gamma(x + 1) - 2\gamma(1 + \lambda)/\lambda];
\end{aligned} \quad (3.10)$$

which is identical to the initial distribution (Eq. 3.9), as required. For lattice-nodes adjacent to the solid-fluid boundaries, the update of some of the populations densities is affected by the boundary nodes (Eq. 3.3); explicitly

$$\begin{aligned}
n'_5(\tfrac{1}{2}) &= [1 - 2\gamma(1/2) + 2\gamma(1 + \lambda)/\lambda], \\
n'_7(\tfrac{1}{2}) &= [1 + 2\gamma(1/2) - 2\gamma(1 + \lambda)/\lambda], \\
n'_6(L - \tfrac{1}{2}) &= [1 + 2\gamma(L - \tfrac{1}{2}) + 2\gamma(1 + \lambda)/\lambda] - 4\gamma L, \\
n'_8(L - \tfrac{1}{2}) &= [1 - 2\gamma(L - \tfrac{1}{2}) - 2\gamma(1 + \lambda)/\lambda] + 4\gamma L;
\end{aligned} \quad (3.11)$$

which, once again, is identical to the initial distribution (Eq. 3.9). Thus the boundary-node collision rules generate an exact linear shear flow; this is because they maintain the second order accuracy of the pure fluid model. Furthermore, the velocity distributions outside and inside the particle are isolated from one another; thus a sharp change in velocity gradient from the inside to the outside the particle surface can be supported, as illustrated in Fig. 4.

As a result of the boundary-node interactions (Eq. 3.3), forces are exerted on the solid particles at the boundary nodes, *i.e.*

$$\begin{aligned}
\mathbf{f}(\mathbf{r} + \tfrac{1}{2}\mathbf{c}_i, t + \tfrac{1}{2}) &= -[n_i(\mathbf{r} + \mathbf{c}_i, t + 1) - n_{i'}(\mathbf{r}, t + 1) - n_i(\mathbf{r}, t_+) + n_{i'}(\mathbf{r} + \mathbf{c}_i, t_+)]\mathbf{c}_i \\
&= 2[n_i(\mathbf{r}, t_+) - n_{i'}(\mathbf{r} + \mathbf{c}_i, t_+) - 2a_1^{c_i} \rho \mathbf{u}_b \cdot \mathbf{c}_i]\mathbf{c}_i;
\end{aligned} \quad (3.12)$$

thus momentum is exchanged locally between the fluid and the solid particle, but the combined momentum of solid and fluid is conserved. The forces and torques on the solid particle are obtained by summing $\mathbf{f}(\mathbf{r} + \frac{1}{2}\mathbf{c}_i)$ and $(\mathbf{r} + \frac{1}{2}\mathbf{c}_i) \times \mathbf{f}(\mathbf{r} + \frac{1}{2}\mathbf{c}_i)$ over all the boundary nodes associated with a particular particle. As an example, Eq. 3.12 can be used to compute the drag force per unit area on a planar wall adjacent to a steadily shearing fluid. We compute the force on one face of each solid boundary surface, assuming that the fluid on the other side is moving uniformly with the velocity of the boundary (as shown in Fig. 4) and therefore exerts no force on the wall; this corresponds to replacing one of the distributions in Eq. 3.12 by its equilibrium form (Eq. 2.39) with a velocity equal to the wall velocity. Using the distributions at $t_+$ just after the molecular collision process (Eq. 3.9 with $1/\lambda$ replaced by $(1 + \lambda)/\lambda$) we find for the wall forces

$$f_y(0) = 2[-n_6(\tfrac{1}{2}, t_+) + n_8(\tfrac{1}{2}, t_+)] = -4(2/\lambda + 1)\gamma = \eta\gamma,$$
$$f_y(L) = 2[n_5(L - \tfrac{1}{2}, t_+) - n_7(L - \tfrac{1}{2}, t_+) - 4\gamma L] = 4(2/\lambda + 1)\gamma = -\eta\gamma; \qquad (3.13)$$

the last equality follows from summing the collisional and lattice contributions to the viscosity (Eq. 2.34), using $\rho = 24$. Thus the wall force is computed exactly for linear shear flows.

As a preliminary application of the method to time-dependent flows, we consider the evolution of the flow field from an impulsively started flat plate. The geometry is similar to Fig. 4, with the plates being sufficiently far apart that they do not interact over the duration of the simulation. We focus on a single plate ($x = 0$). Initially the system is at rest; at $t = 0$, an impulsive force gives the plate a constant velocity $[0, U, 0]$. In this problem it is again convenient to define the lattice nodes at half-integer values of $x$, and for the fluid to reside at the lattice nodes at half-integer values of the time; then the boundary conditions at the plate are applied at $x = 0$ and $t = 0$ precisely. We compute numerically the evolution of the flow field $[0, u(x, t), 0]$, created by the diffusion of vorticity into the fluid, and compare with the analytic solutions for the velocity field (Batchelor, 1967)

$$u(x, t) = U\left[1 - \Phi(x/\sqrt{4\nu t})\right], \qquad (3.14)$$

and the force per unit area

$$f(t) = -\eta \nabla_x u(0,t) = -\eta U(\pi \nu t)^{-1/2}; \qquad (3.15)$$

$\nu = \eta/\rho$ is the kinematic viscosity of the fluid and $\Phi$ is the error function. The results are shown in Fig. 5 (for a viscosity $\nu = 1/6$) at several different times. It can be seen that there is complete agreement, except at very short times and distances. This simple test implies that we can accurately simulate both stationary and time-dependent flows, as will be confirmed by further results in paper II.

Next we consider flow perpendicular to the wall. The incompressibility condition means that there can be no velocity gradients in steady flow; thus the fluid velocity and the wall velocities are $[u, 0, 0]$. For this flow we can further simplify the model in Fig. 4 to only three velocity directions; stationary particles, and particles moving in the positive and negative $x$ directions. The projected distributions are

$$n_0(x) = 12,$$
$$n_1(x) = 6(1 + 2u), \; n_2(x) = 6(1 - 2u), \qquad (3.16)$$

Clearly this distribution is stationary with the respect to the evolution of the lattice-Boltzmann equation; it is also stationary with respect to the boundary-node update rules (Eq. 3.3). However this is not the only stationary distribution that satisfies the boundary conditions. A more general form for the distribution function, which exhibits a two time-step repeat cycle is

$$n_0(x) = 12,$$
$$n_1(x) = 6(1 + 2u + (-1)^{t+x} 2w), \; n_2(x) = 6(1 - 2u - (-1)^{t+x} 2w); \qquad (3.17)$$

it is straightforward to verify that this distribution also satisfies both the time evolution equation and the boundary conditions at the walls. Thus the fluid has a uniform momentum $\rho u$ and a "staggered" momentum $(-1)^{t+x} \rho w$. Staggered momenta are an artifact of all lattice models (Zanetti, 1989); the precise value of the staggered momentum depends on the channel width (in this example) and the initial conditions. Although staggered momenta cannot arise spontaneously in the fluid, they can be generated at solid surfaces, as we see in this example. However, it can be shown, by techniques similar to those used above, that staggered momentum parallel to the walls is damped by the boundary conditions, so that for the plane Couette flow problem, the steady solutions have no staggered momentum component.

In the more complex geometries that are of interest in particulate suspensions, it is impossible to analyze the staggered momenta analytically. Numerical results show that large oscillations in particle torques can be built up by a feedback mechanism in which the staggered momenta are fed by ever increasing angular velocities of the particles. To overcome these instabilities we average the force and fluid velocity over two successive time steps which effectively cancels out the staggered momentum contribution. In the above example, this gives a uniform flow field with velocity $[u, 0, 0]$ regardless of the magnitude of the staggered momenta. Since the forces at the boundary nodes are generated at the half-integer time steps, the smoothly varying force $\bar{\mathbf{f}}$ at the intermediate integer time is

$$\bar{\mathbf{f}}(\mathbf{r} + \tfrac{1}{2}\mathbf{c}_i, t) = (1/2)[\mathbf{f}(\mathbf{r} + \tfrac{1}{2}\mathbf{c}_i, t - \tfrac{1}{2}) + \mathbf{f}(\mathbf{r} + \tfrac{1}{2}\mathbf{c}_i, t + \tfrac{1}{2})]. \tag{3.18}$$

We can calculate the smooth part of the fluid velocity field at half-integer time steps in a similar way

$$\bar{\mathbf{u}}(\mathbf{r}, t + \tfrac{1}{2}) = (1/2)[\mathbf{u}(\mathbf{r}, t) + \mathbf{u}(\mathbf{r}, t + 1)], \tag{3.19}$$

or at integer time steps using the 3-point formula

$$\bar{\mathbf{u}}(\mathbf{r}, t) = (1/4)[\mathbf{u}(\mathbf{r}, t - 1) + 2\mathbf{u}(\mathbf{r}, t) + \mathbf{u}(\mathbf{r}, t + 1)], \tag{3.20}$$

The velocities of finite mass particles (as opposed to infinitely massive fixed objects) are updated every two time steps,

$$\mathbf{U}(t + 1) = \mathbf{U}(t - 1) + 2M^{-1}\mathbf{F}(t), \quad \mathbf{\Omega}(t + 1) = \mathbf{\Omega}(t - 1) + 2\mathbf{I}^{-1} \cdot \mathbf{T}(t). \tag{3.21}$$

The particle mass $M$ and moment of inertia $\mathbf{I}$ are preassigned parameters which control the rate at which particles respond to the fluid flow; usually $M$ and $\mathbf{I}$ are on the order of several thousands (in lattice units). Since the velocities vary slowly on the time scale of a lattice-Boltzmann cycle, the precise form for the update is usually not too important; however, it is important to use time-smoothed forces and torques, as described in Eq. 3.18.

So far we have considered only linear shear flows. In the next and last example we consider stationary two-dimensional channel or Poiseille flow. The geometry is, once again, as shown in Fig. 4, with the walls at rest ($\mathbf{u}_b = 0$). The fluid is driven by a pressure gradient, which is represented in the simulation by a uniform force density in the fluid. Thus we apply a constant increment $\Delta j_y$ to the $y$ momentum at each node so that the pressure gradient down the channel is $\nabla_y p = \Delta j_y$. The fluid velocity at each node is measured after half the force has been applied; we have found empirically that this prescription gives the fastest convergence as a function of system size. The steady flow profiles are compared with the analytic solution,

$$u_y(x) = (\Delta j_y/2\eta)\frac{x}{L}\left(1 - \frac{x}{L}\right), \qquad (3.22)$$

in Fig. 6. It can be seen that the agreement is very good for channels more than about 9 lattice spacings wide. Furthermore, we find that the force on the wall is exactly correct at steady state, no matter what the channel width. This result follows from the balance between forces on the walls and the total force from the pressure gradient. Since the pressure forces are distributed equally on the walls it follows that the wall force per unit area $f = \Delta j_y L/2$, which is the correct result for Poiseille flow.

In this section we have seen how moving solid boundaries can be incorporated into a lattice-Boltzmann simulation, and we have indicated how they function for a few simple examples. In paper II we describe numerical results for spherical particles, both periodic arrays and random assemblies. Results are compared with known analytic and numerical solutions of the creeping-flow and Navier-Stokes equations.

## IV. FLUCTUATIONS

In recent years, it has become increasingly obvious that the the lattice-Boltzmann equation is a much better simulation tool for hydrodynamics than lattice gases. However, in its normal state the lattice-Boltzmann equation cannot model the molecular fluctuations in the solvent that give rise to Brownian motion. Of course in many situations Brownian motion is unimportant, but, for suspensions of sub-micron sized particles, it is a fundamental component of the dynamics; It has been recently shown (Ladd, 1993b) that fluctuations can be incorporated into the lattice-Boltzmann equation, within the framework of fluctuating hydrodynamics (Landau and Lifshitz, 1959), by adding a random component to the fluid stress tensor. Numerical tests showed that the resulting particle motion, in dilute to concentrated suspensions, closely matches recent experimental results (Zhu et al., 1992; Kao et al., 1993), even at very short times where particle inertia plays an important role. In this section, we describe the basic theory of fluctuations as it applies to the lattice-Boltzmann model; in paper II numerical tests of the method for particulate suspensions of spheres will be reported.

The basic idea of fluctuating hydrodynamics is that, on length scales and time scales intermediate between the molecular and the hydrodynamic, thermally-induced fluctuations can be reduced to random fluctuations in the fluxes of the conserved variables; *i.e.* the stress tensor and perhaps the heat flux also. Since the fluxes are included explicitly in a lattice-Boltzmann simulation, it is plausible that molecular fluctuations can be modeled realistically on intermediate scales, even though the details of the microscopic interactions are different. In the present context, this means that the time evolution of the velocity distribution includes a stochastic term $n'_i(\mathbf{r}, t)$, representing the thermally-induced fluctuations in the stress tensor,

$$n_i(\mathbf{r} + \mathbf{c}_i, t) = n_i(\mathbf{r}, t) + \Delta_i(\mathbf{r}, t) + n'_i(\mathbf{r}, t), \tag{4.1}$$

where $n'$ is chosen so that only its stress moment is non zero (*c.f.* Eq. 3.4),

$$n'_i = -a_2^{c_i} \sigma'_{\alpha\beta} \overline{c_{i\alpha} c_{i\beta}}. \tag{4.2}$$

These random stress fluctuations are uncorrelated in space and time (Landau and Lifshitz, 1959) and are sampled from a Gaussian distribution,

$$\left\langle \sigma'_{\alpha\beta}(\mathbf{r},t) \sigma'_{\gamma\delta}(\mathbf{r}',t') \right\rangle = A \delta_{\mathbf{rr}'} \delta_{tt'} \left( \delta_{\alpha\gamma}\delta_{\beta\delta} + \delta_{\alpha\delta}\delta_{\beta\gamma} - (2/3)\delta_{\alpha\beta}\delta_{\gamma\delta} \right); \tag{4.3}$$

the choice of the variance $A$ serves to define the effective temperature of the fluid. We can determine the relationship between $A$ and the temperature via the fluctuation-dissipation theorem.

It is convenient in what follows to work in Fourier space, defining the Fourier transform of the velocity distribution function as

$$n_i(\mathbf{k},t) = \sum_{\mathbf{r} \in V} e^{-i\mathbf{k}\cdot\mathbf{r}} n_i(\mathbf{r},t), \tag{4.4}$$

where the sum is over all the lattice points in the periodic unit cell. Then the equation for momentum conservation (Eq. 2.17) can be written as

$$\sum_i [e^{i\mathbf{k}\cdot\mathbf{c}_i} n_i(\mathbf{k},t+1) - n_i(\mathbf{k},t)] \mathbf{c}_i = 0; \tag{4.5}$$

we will attempt to cast this equation into the form of a Langevin equation for the transverse (or solenoidal) momentum fluctuations $\mathbf{j}_\perp(\mathbf{k},t) = (\mathbf{1} - \hat{\mathbf{k}}\hat{\mathbf{k}}) \cdot \mathbf{j}(\mathbf{k},t)$ ($\hat{\mathbf{k}}$ is the unit vector $\mathbf{k}/k$). Expanding the exponential, we have

$$\mathbf{j}_\perp(\mathbf{k},t+1) - \mathbf{j}_\perp(\mathbf{k},t) + i\mathbf{k} \cdot \overline{\overline{\Pi}}(\mathbf{k},t+1) \cdot (\mathbf{1} - \hat{\mathbf{k}}\hat{\mathbf{k}}) - (k^2/6) \mathbf{j}_\perp(\mathbf{k},t+1) = \mathcal{O}(k^4), \tag{4.6}$$

where $\overline{\overline{\Pi}}(\mathbf{k},t) = \sum_i n_i(\mathbf{k},t) \overline{\mathbf{c}_i \mathbf{c}_i}$ is the traceless part of the momentum flux. To obtain a Langevin equation, we must divide the momentum flux into a slowly relaxing dissipative part and a rapidly varying fluctuating part. In section II, it was shown that on the shorter $\epsilon^{-1}$ time scale, the stress relaxed to its Navier-Stokes form (Eq. 2.30), whereas the velocity gradients varied on a longer $\epsilon^{-2}$ time scale. Thus we define the fluctuating stress $\mathbf{\Sigma}$ as

$$\Sigma_{\alpha\beta} = -[\overline{\Pi}_{\alpha\beta} - (1/3\lambda)(\nabla_\alpha j_\beta + \nabla_\beta j_\alpha - (2/3)\nabla_\gamma j_\gamma \delta_{\alpha\beta}], \tag{4.7}$$

where we have ignored density fluctuations (because they do not couple to the transverse momentum flux) and non-linear $\rho\mathbf{uu}$ terms (because we are examining small length scales where the Reynolds number is negligible). Combining Eqs. 4.6 and 4.7, and summing the collisional $(1/3\lambda)$ and lattice $(1/6)$ contributions to the viscosity (Eq. 2.34), we obtain a discrete Langevin equation for the transverse momentum fluctuations,

$$\mathbf{j}_\perp(\mathbf{k}, t+1) - \mathbf{j}_\perp(\mathbf{k}, t) + \nu k^2 \mathbf{j}_\perp(\mathbf{k}, t+1) = \imath \mathbf{k} \cdot \boldsymbol{\Sigma}(\mathbf{k}, t+1) \cdot (\mathbf{1} - \hat{\mathbf{k}}\hat{\mathbf{k}}), \quad (4.8)$$

with a random force $\imath \mathbf{k} \cdot \boldsymbol{\Sigma}(\mathbf{k}, t+1) \cdot (\mathbf{1} - \hat{\mathbf{k}}\hat{\mathbf{k}})$. The solution of a discrete Langevin equation is described in the Appendix; the result in this case is (Eq. A9)

$$2\nu \left\langle \mathbf{j}_\perp(\mathbf{k}) \cdot \mathbf{j}_\perp(-\mathbf{k}) \right\rangle = \sum_{t=-\infty}^{\infty} \left\langle \hat{\mathbf{k}} \cdot \boldsymbol{\Sigma}(\mathbf{k}, t) \cdot (\mathbf{1} - \hat{\mathbf{k}}\hat{\mathbf{k}}) \cdot \boldsymbol{\Sigma}(-\mathbf{k}, 0) \cdot \hat{\mathbf{k}} \right\rangle. \quad (4.9)$$

The effective temperature of the fluid can be defined by equating the fluctuations in momentum to those in real fluids; for a molecular fluid of $N$ particles of mass $m$,

$$\left\langle \mathbf{j}_\perp(\mathbf{k}) \cdot \mathbf{j}_\perp(-\mathbf{k}) \right\rangle = \frac{2}{3} \left\langle \sum_{i,j=1}^{N} e^{-\imath \mathbf{k} \cdot \mathbf{R}_{ij}} \mathbf{v}_i \cdot \mathbf{v}_j \right\rangle = 2Nmk_BT = 2\rho V k_BT, \quad (4.10)$$

where the factor 2/3 comes from the missing longitudinal fluctuations. In the long-wavelength limit we can isotropically average over the directions of $\mathbf{k}$ with the result

$$\lim_{\mathbf{k} \to 0} \left\langle \hat{\mathbf{k}} \cdot \boldsymbol{\Sigma}(\mathbf{k}, t) \cdot (\mathbf{1} - \hat{\mathbf{k}}\hat{\mathbf{k}}) \cdot \boldsymbol{\Sigma}(-\mathbf{k}, 0) \cdot \hat{\mathbf{k}} \right\rangle = \frac{1}{5} \left\langle \boldsymbol{\Sigma}(t) : \boldsymbol{\Sigma}(0) \right\rangle = 2 \left\langle \Sigma_{xy}(t) \Sigma_{xy}(0) \right\rangle, \quad (4.11)$$

where

$$\boldsymbol{\Sigma}(t) = \lim_{\mathbf{k} \to 0} \boldsymbol{\Sigma}(\mathbf{k}, t) = \sum_{\mathbf{r} \in V} \boldsymbol{\sigma}(\mathbf{r}, t). \quad (4.12)$$

Combining Eqs. 4.9–4.11 we obtain a fluctuation formula for the viscosity

$$\eta V k_BT = (1/2) \left\langle \Sigma_{xy}(0)\Sigma_{xy}(0) \right\rangle + \sum_{t=1}^{\infty} \left\langle \Sigma_{xy}(t)\Sigma_{xy}(0) \right\rangle, \quad (4.13)$$

which is an analogue of the Green-Kubo relation for molecular liquids (Hansen and McDonald, 1986), discretized in time. The summation in Eq. 4.13 is the equivalent of a Simpson's rule approximation to the time integral that appears in the usual Green-Kubo formulae; in paper II this expression is used to calculate suspension viscosities, replacing the fluid stress tensor with the combined stress tensor of the solid and fluid phases.

Let us now consider the time evolution of the stress tensor in the pure fluid. The total stress fluctuations $\boldsymbol{\Sigma}$ in a volume $V$ are independent of the propagation of population density; they vary only because of collisions and random fluctuations. Thus the time evolution of $\boldsymbol{\Sigma}$ including the random fluctuations can be written as

$$\boldsymbol{\Sigma}(t+1) = (1+\lambda)\boldsymbol{\Sigma}(t) + \sum_{\mathbf{r} \in V} \boldsymbol{\sigma}'(\mathbf{r}, t). \tag{4.14}$$

Since $\langle \boldsymbol{\sigma}'(\mathbf{r}, t) \boldsymbol{\Sigma}(0) \rangle = 0$ for all $t \geq 0$, the time correlation function of the stress fluctuations can be written in terms of the equal time correlation function (c.f. Eq. A4),

$$\langle \boldsymbol{\Sigma}(t) \boldsymbol{\Sigma}(0) \rangle = (1+\lambda)^t \langle \boldsymbol{\Sigma}(0) \boldsymbol{\Sigma}(0) \rangle ; \tag{4.15}$$

thus from Eqs. 4.13 and 2.34 we can relate the effective temperature to the equal-time stress fluctuations,

$$\rho V k_B T = 3 \left\langle \Sigma_{xy}^2 \right\rangle. \tag{4.16}$$

Lastly, we need to relate the equal-time stress fluctuations to the fluctuations in the random stress tensor. This can be done by noting that Eq. 4.14 is a discretized Langevin equation for the stress tensor, in which the random forces at different times are uncorrelated (see Appendix). Since $| (1+\lambda) | < 1$, the equal-time correlation function is given by (Eq. A8)

$$\langle \Sigma_{\alpha\beta} \Sigma_{\gamma\delta} \rangle = \frac{AV}{1 - (1+\lambda)^2} \left( \delta_{\alpha\gamma}\delta_{\beta\delta} + \delta_{\alpha\delta}\delta_{\beta\gamma} - (2/3)\delta_{\alpha\beta}\delta_{\gamma\delta} \right). \tag{4.17}$$

Then, from Eqs. 4.16 and 4.17

$$A = (\rho k_B T/3)[1 - (1+\lambda)^2] = 2\eta k_B T \lambda^2, \tag{4.18}$$

which is the fluctuation-dissipation relation for our fluctuating lattice-Boltzmann equation. It defines the effective temperature of the fluid so that the dissipative (Eq. 2.34) and fluctuating (Eq. 4.13) expressions for the viscosity are consistent.

Some numerical tests of the fluctuating lattice-Boltzmann equation are illustrated in Fig. 7. Here we compare simulation results for the stress-stress correlation function with theoretical results (Eq. 4.15), for a range of values of $\lambda$ covering kinematic viscosities from about $10^{-3}$ to about 30. For both the over-relaxing collision operators $(-2 < \lambda < -1)$, where the stress tensor changes sign at every time step, and for the under-relaxing collision operators $(-1 < \lambda < 0)$ the agreement between theory and simulation is perfect. In general, we will choose values of $\lambda$ close to $-1$, to minimize the relaxation time of the stress fluctuations. The special case $\lambda = -1$ is particularly useful; here the stress fluctuations decay instantaneously, since only the random part of the stress tensor gets propagated at each time step (see Eq. 4.14). In this limit we find that the expression for $A$ (Eq. 4.18) reduces to the Landau-Lifshitz result (Landau and Lifshitz, 1959) $A = 2\eta k_B T$.

## V. CONCLUSIONS

In this paper the theoretical foundation for a new simulation technique for particulate suspensions has been described. Much of the theory underlying this work is now well understood; in particular the macroscopic fluid dynamics arising from the various lattice-Boltzmann models, and the modeling of hydrodynamic stick boundary conditions by local modifications to particle populations. The computational effectiveness of the method will be demonstrated in paper II. The inclusion of fluctuations in the stress tensor is a more recent development (Ladd, 1993b) and is less well understood. Nevertheless, we have been able to derive discrete analogues of the basic equations of fluctuating hydrodynamics; again the numerical tests reported in paper II provide convincing numerical evidence of the correctness of the approach.

## ACKNOWLEDGMENTS

This work was supported by the U.S. Department of Energy and Lawrence Livermore National Laboratory under Contract No. W-7405-Eng-48.

FIGURES

Fig. 1. Microscopic models of a colloidal suspension. The dynamical properties of the large particles are insensitive to the detailed motions of the background fluid, so that the continuum fluid in Fig. 1a can be equally well replaced by either a molecular solvent (Fig. 1b) or a lattice gas (Fig. 1c). Lattice-gas/lattice-Boltzmann simulations are many orders of magnitude faster computationally.

Fig. 2. Location of the boundary nodes for a circular object of radius 2.5 lattice spacings. The velocities along links cutting the boundary surface are indicated by arrows. The location of the boundary nodes are shown by solid squares and the lattice nodes by solid circles.

Fig. 3. Population densities before and after a collision with a boundary node. The effects of stationary (Fig. 3a) and moving (Figs. 3b and 3c) boundary nodes on the incoming populations are shown. The arrows indicate the velocity direction and the lengths of the solid lines are proportional to the population densities. The differences between population densities are highly exaggerated for clarity. Note that the effects of the moving boundary are the same in Figs 3b and 3c, because the velocity component parallel to the link direction is the same.

Fig. 4. A two-dimensional projection of the lattice-Boltzmann fluid, bounded by plane walls. The boundary nodes are shown as squares and the boundary planes ($x = 0$ and $x = L$) by dashed lines. The circles are the lattice nodes; the set of nodes explicitly considered in the text are shown filled. The fluid between the walls is subjected to a velocity gradient by the relative motion of the walls; the fluid outside the walls moves with uniform velocity equal to the wall velocity. The labeling of velocity directions used in the text is also shown; velocity components in the $z$-direction have been projected onto the $xy$-plane. There is an additional density of stationary particles (not shown), labeled 0, corresponding to velocities $[0, 0, \pm 1]$. The lower portion of the figure is a plot of the velocity profile, with the crosses showing the fluid velocity $u_y$ at the nodes and the dotted line the interpolation between nodes.

Fig. 5. Flow induced by an impulsively loaded flat plate, $U(t) = 0$ for $t < 0$ and $U(t) = U$ for $t \geq 0$. The plots show the velocity field and force per unit area on the wall at various times; the solid circles are the numerical simulations, and the solid lines are the analytical results (Batchelor, 1967).

Fig. 6. Poiseuille flow in two-dimensional channels for different channel widths, L. The simulated velocities (solid circles) are scaled by the theoretical maximum value $u_0 = \Delta j_y L^2 / 8\eta$.

Fig. 7. Stress-stress correlation function for a lattice-Boltzmann fluid. Numerical results (solid circles) and theoretical results from Eq. 4.15 (solid lines) are shown for different values of $\lambda$.

# APPENDIX: DISCRETE LANGEVIN EQUATION

The discrete Langevin equation (Eqs. 4.8 or 4.14) is of the general form

$$j(t+1) - j(t) = -\alpha j(t) + f(t), \tag{A1}$$

where $\alpha$ is a positive constant controlling the rate of dissipation ($0 < \alpha < 2$), and $f(t)$ is the random force. The random force has the usual property

$$\langle f(t) j(t') \rangle = 0, \tag{A2}$$

for all $t \geq t'$. We can rewrite Eq. A1 as

$$j(t+t') = (1-\alpha)^{t'} j(t) + \sum_{s=1}^{t'} (1-\alpha)^{s-1} f(t+t'-s), \tag{A3}$$

and from Eq. A2 we can express the time correlation function in terms of equal-time fluctuations;

$$\langle j(t+t') j(t) \rangle = (1-\alpha)^{t'} \langle j(t) j(t) \rangle. \tag{A4}$$

The equal time correlations, measured from some initial time ($t = 0$), are given by

$$\langle j(t') j(t') \rangle = (1-\alpha)^{2t'} \langle j(0) j(0) \rangle + \sum_{s=1}^{t'} \sum_{s'=1}^{t'} (1-\alpha)^{s+s'-2} \langle f(t'-s) f(t'-s') \rangle. \tag{A5}$$

In the long-time ($t' \to \infty$) limit the system loses all memory of its initial conditions; the equal-time fluctuations can then be written as

$$\langle j^2 \rangle = \sum_{s=0}^{\infty} \sum_{s'=0}^{\infty} (1-\alpha)^{s+s'} \langle f(s) f(s') \rangle; \tag{A6}$$

Equation A6 can be simplified by a change of variables, $s_\pm = s \pm s'$;

$$\langle j^2 \rangle = \sum_{s_-=-\infty}^{\infty} (1-\alpha)^{|s_-|} \langle f(s_-) f(0) \rangle \sum_{s_+=0}^{\infty} (1-\alpha)^{2s_+}$$
$$= \frac{1}{2\alpha - \alpha^2} \sum_{s_-=-\infty}^{\infty} (1-\alpha)^{|s_-|} \langle f(s_-) f(0) \rangle. \tag{A7}$$

In the special case that the random force is delta-function correlated in time (as in Eq. 4.14), then

$$\left\langle j^2 \right\rangle = \frac{\langle f^2 \rangle}{2\alpha - \alpha^2}. \tag{A8}$$

In the more general case (*i.e.* in Eq. 4.8), the random force has a finite, although short, relaxation time. Here we can only obtain a simple result for small values of $\alpha$, corresponding to the $\mathbf{k} \to 0$ limit in Eq. 4.8;

$$\left\langle j^2 \right\rangle = \frac{1}{2\alpha} \sum_{t=-\infty}^{\infty} \langle f(t)f(0) \rangle, \tag{A9}$$

which agrees with Eq. A8 when $\alpha$ is small and the random force is delta-function correlated.